\documentclass[10pt, conference, letterpaper]{IEEEtran}
\usepackage{url}
\usepackage[utf8]{inputenc}
\usepackage[english]{babel}
\usepackage{xcolor}
\usepackage{amsmath}
\usepackage{amssymb}
\usepackage{bm}      

\usepackage{xargs}

\usepackage[acronyms,nonumberlist,nopostdot,nomain,nogroupskip]{glossaries}
\usepackage{tablefootnote}
\usepackage{booktabs}
\usepackage{tabularx}
\usepackage{epsfig}
\usepackage[outdir=./]{epstopdf}
\usepackage{tikz}
\usepackage{pgfplots}
\pgfplotsset{compat=newest}
\pgfplotsset{plot coordinates/math parser=false}
\newlength\fheight
\newlength\fwidth
\usetikzlibrary{plotmarks,shapes,patterns,decorations.pathreplacing,backgrounds,calc,arrows,arrows.meta,spy,matrix}
\usepgfplotslibrary{patchplots,groupplots}
\usepackage{tikzscale}
\usepackage{physics}
\usepackage{pgfkeys}
\usepackage{multirow}
\usepackage[font=scriptsize]{subcaption}
\usepackage[font=footnotesize]{caption}

\usepackage{mathtools}
\usepackage{xfrac}

\usepackage{dblfloatfix}    
\usepackage{xcolor,colortbl}
\definecolor{Gray}{gray}{0.85}
\definecolor{LightCyan}{rgb}{0.88,1,1}

\usepackage[per-mode=fraction,binary-units]{siunitx}
\DeclareSIUnit\RPM{\text{RPM}}
\DeclareSIUnit\point{\text{point}}
\DeclareSIUnit\frame{\text{frame}}
\DeclareSIUnit\fps{\text{fps}}

\makeglossaries
\newglossaryentry{gpcc}
{
    name=G-PCC,
    description={Geometry-based Point Cloud Compression}
}
\newglossaryentry{png}
{
    name=PNG,
    description={Portable Network Graphics}
}

\newglossaryentry{tof}
{
    name=ToF,
    description={Time-of-Flight}
}

\newglossaryentry{LiDAR}
{
    name=LiDAR,
    description={Light Detection and Ranging \emph{or} Laser Imaging Detection and Ranging}
}

\newglossaryentry{Deflate}
{
    name=DEFLATE,
    description={Lossless data compression file format that uses a combination of \gls{LZW} and Huffman coding}
}

\newglossaryentry{inter-frame}
{
    name=inter-frame,
    description={It is a frame in a video compression stream which is expressed in terms of one or more neighboring frames (through prediction)}
}

\newglossaryentry{intra-frame}
{
    name=intra-frame,
    description={Video compression technique that does not involved adjacent frames}
}

\newglossaryentry{point-cloud}
{
    name=point cloud,
    description={A set of individual points in a 3-Dimensional space}
}

\newglossaryentry{lossless}
{
    name=lossless,
    description={It is a class of data compression algorithms that allows the original data to be perfectly reconstructed from the compressed data. By contrast, lossy compression permits reconstruction only of an approximation of the original data, though usually with greatly improved compression rates}
}

\newglossaryentry{octree}
{
    name=Octree,
    description={A tree (data structure) where each internal node has exactly eight children}
}

\newglossaryentry{voxel}
{
    name=voxel,
    description={Volumetric Picture Element, the equivalent of a 2D-image pixel but on a regular grid in a 3-Dimensional space}
}

\newglossaryentry{image-based}
{
    name=image-based,
    description={Compression method that implements image formats}
}

\newglossaryentry{video-based}
{
    name=video-based,
    description={Compression method that implements video formats}
}

\newglossaryentry{dictionary}
{
    name=dictionary,
    description={A set of strings contained in a data structure maintained by the encoder. When the encoder ﬁnds a match, it substitutes a reference to the string’s position in the data structure}
}

\newacronym{FOV}{FOV}{field of view}
\newacronym{GPS}{GPS}{Global Positioning System}
\newacronym{Bpp}{BPP}{Bytes per Point}
\newacronym{PSNR}{PSNR}{Peak Signal to Noise Ratio}
\newacronym{SNR}{SNR}{Signal to Noise Ratio}
\newacronym{MSE}{MSE}{Mean Square Error}
\newacronym{PCL}{PCL}{Point Cloud Library}
\newacronym{pcd}{PCD}{Point Cloud Data}
\newacronym{PNG}{PNG}{Portable Network Graphics}
\newacronym{TIFF}{TIFF}{Tagged Image File Format}
\newacronym{PPM}{PPM}{Portable Pixmap Format}
\newacronym{J-LS}{J-LS}{Lossless JPEG}
\newacronym{ZIP}{ZIP}{Compression using Deflate algorithm}
\newacronym{RGB}{RGB}{Red-Green-Blue}
\newacronym{AVI}{AVI}{Audio Video Interleave}
\newacronym{MPEG}{MPEG}{Moving Picture Experts Group}
\newacronym{MPEG-4}{MPEG-4}{Moving Picture Experts Group multimedia container part 4 or MP4}
\newacronym{MJ2}{MJ2}{Motion JPEG 2000}
\newacronym{H264}{H264}{Advanced Video Coding}
\newacronym{LZW}{LZW}{Lempel–Ziv–Welch}
\newacronym{XYZ}{XYZ}{Cartesian coordinates}
\newacronym{CSV}{CSV}{Comma-separated Values (Microsoft © Excel)}
\newacronym{PCAP}{PCAP}{Packet Capture}
\newacronym{NaN}{NaN}{Not a Number}
\newacronym{5G}{5G}{fifth generation}
\newacronym{VLC}{VLC}{Variable Length Code}
\newacronym{LAS}{LAS}{LASer (binary format for archiving point-cloud data)}
\newacronym{DCT}{DCT}{Discrete Cosine Transform}
\newacronym{3gpp}{3GPP}{3rd Generation Partnership Project}
\newacronym{adc}{ADC}{Analog to Digital Converter}
\newacronym{5g}{5G}{5th generation}
\newacronym{aimd}{AIMD}{Additive Increase Multiplicative Decrease}
\newacronym{am}{AM}{Acknowledged Mode}
\newacronym{amc}{AMC}{Adaptive Modulation and Coding}
\newacronym{aqm}{AQM}{Active Queue Management}
\newacronym{awgn}{AGWN}{Additive White Gaussian Noise}
\newacronym{balia}{BALIA}{Balanced Link Adaptation}
\newacronym{bdp}{BDP}{Bandwidth-Delay Product}
\newacronym{bf}{BF}{beamforming}
\newacronym{cc}{CC}{Congestion Control}
\newacronym{cdf}{CDF}{Cumulative Distribution Function}
\newacronym{cn}{CN}{Core Network}
\newacronym{cqi}{CQI}{Channel Quality Information}
\newacronym{cp}{CP}{Control Plane}
\newacronym{csirs}{CSI-RS}{Channel State Information - Reference Signal}
\newacronym{dc}{DC}{Dual Connectivity}
\newacronym{rb}{RB}{Resource Block}
\newacronym{dce}{DCE}{Direct Code Execution}
\newacronym{dci}{DCI}{Downlink Control Information}
\newacronym{udp}{UDP}{User Datagram Protocol}
\newacronym{dl}{DL}{Downlink}
\newacronym{dmr}{DMR}{Deadline Miss Ratio}
\newacronym{dmrs}{DMRS}{DeModulation Reference Signal}
\newacronym{e2e}{E2E}{End-to-End}
\newacronym{ppp}{PPP}{Poission Point Process}
\newacronym{si}{SI}{Study Item}
\newacronym{ecn}{ECN}{Explicit Congestion Notification}
\newacronym{edf}{EDF}{Earliest Deadline First}
\newacronym{enb}{eNB}{eNodeB}
\newacronym{epc}{EPC}{Evolved Packet Core}
\newacronym{es}{ES}{Edge Server}
\newacronym{cav}{CAV}{connected and autonomous vehicle}
\newacronym{fdma}{FDMA}{Frequency Division Multiple Access}
\newacronym{fdd}{FDD}{Frequency Division Duplexing}
\newacronym{upa}{UPA}{Uniform Planar Array}
\newacronym[firstplural=Radio Access Technologies (RATs)]{rat}{RAT}{Radio Access Technology}
\newacronym[firstplural=Radio Access Technology (RTs)]{rt}{RT}{Radio Technology}
\newacronym{fs}{FS}{Fast Switching}
\newacronym{isd}{ISD}{inter-site distance}
\newacronym{ftp}{FTP}{File Transfer Protocol}
\newacronym{gnb}{gNB}{Next Generation Node Base}
\newacronym{harq}{HARQ}{Hybrid Automatic Repeat reQuest}
\newacronym{hetnet}{HetNet}{Heterogeneous Network}
\newacronym{hh}{HH}{Hard Handover}
\newacronym{hol}{HOL}{Head-of-Line}
\newacronym{ia}{IA}{Initial Access}
\newacronym{imt}{IMT}{International Mobile Telecommunication}
\newacronym{iot}{IoT}{Internet of Things}
\newacronym{kpi}{KPI}{Key Performance Indicator}
\newacronym{los}{LOS}{Line of Sight}
\newacronym{lte}{LTE}{Long Term Evolution}
\newacronym{m2m}{M2M}{Machine to Machine}
\newacronym{mac}{MAC}{Medium Access Control}
\newacronym{mc}{MC}{Multi-Connectivity}
\newacronym{mcs}{MCS}{Modulation and Coding Scheme}
\newacronym{mec}{MEC}{Mobile Edge Cloud}
\newacronym{mi}{MI}{Mutual Information}
\newacronym{mimo}{MIMO}{Multiple Input Multiple Output}
\newacronym{mmwave}{mmWave}{millimeter wave}
\newacronym{mptcp}{MPTCP}{Multipath TCP}
\newacronym{mr}{MR}{Maximum Rate}
\newacronym{mss}{MSS}{Maximum Segment Size}
\newacronym{mtd}{MTD}{Machine-Type Device}
\newacronym{mtu}{MTU}{Maximum Transmission Unit}
\newacronym{nfv}{NFV}{Network Function Virtualization}
\newacronym{nlos}{NLOS}{Non Line of Sight}
\newacronym{nlosb}{NLOSb}{Building Non Line of Sight}
\newacronym{nlosv}{NLOSv}{Vehicle Non Line of Sight}
\newacronym{nr}{NR}{New Radio}
\newacronym{ofdm}{OFDM}{Orthogonal Frequency Division Multiplexing}
\newacronym{pdcch}{PDCCH}{Physical Downlonk Control Channel}
\newacronym{pdcp}{PDCP}{Packet Data Convergence Protocol}
\newacronym{pdsch}{PDSCH}{Physical Downlink Shared Channel}
\newacronym{pdu}{PDU}{Packet Data Unit}
\newacronym{pf}{PF}{Proportional Fair}
\newacronym{pgw}{PGW}{Packet Gateway}
\newacronym{phy}{PHY}{Physical}
\newacronym{pbch}{PBCH}{Physical Broadcast Channel}
\newacronym[plural=\gls{mme}s,firstplural=Mobility Management Entities (MMEs)]{mme}{MME}{Mobility Management Entity}
\newacronym{prb}{PRB}{Physical Resource Block}
\newacronym{pss}{PSS}{Primary Synchronization Signal}
\newacronym{pucch}{PUCCH}{Physical Uplink Control Channel}
\newacronym{pusch}{PUSCH}{Physical Uplink Shared Channel}
\newacronym{rach}{RACH}{Random Access Channel}
\newacronym{ran}{RAN}{Radio Access Network}
\newacronym{red}{RED}{Random Early Detection}
\newacronym{rf}{RF}{Radio Frequency}
\newacronym{rlc}{RLC}{Radio Link Control}
\newacronym{rlf}{RLF}{Radio Link Failure}
\newacronym{rrc}{RRC}{Radio Resource Control}
\newacronym{rrm}{RRM}{Radio Resource Management}
\newacronym{rr}{RR}{Round Robin}
\newacronym{rs}{RS}{Remote Server}
\newacronym{rsrp}{RSRP}{Reference Signal Received Power}
\newacronym{rss}{RSS}{Received Signal Strength}
\newacronym{rtt}{RTT}{Round Trip Time}
\newacronym{rw}{RW}{Receive Window}
\newacronym{rx}{RX}{Receiver}
\newacronym{sa}{SA}{standalone}
\newacronym{sack}{SACK}{Selective Acknowledgment}
\newacronym{sap}{SAP}{Service Access Point}
\newacronym{sch}{SCH}{Secondary Cell Handover}
\newacronym{scoot}{SCOOT}{Split Cycle Offset Optimization Technique}
\newacronym{sdma}{SDMA}{Spatial Division Multiple Access}
\newacronym{sinr}{SINR}{Signal to Interference plus Noise Ratio}
\newacronym{sm}{SM}{Saturation Mode}
\newacronym{snr}{SNR}{Signal to Noise Ratio}
\newacronym{son}{SON}{Self-Organizing Network}
\newacronym{ss}{SS}{Synchronization Signal}
\newacronym{srs}{SRS}{Sounding Reference Signal}
\newacronym{sss}{SSS}{Secondary Synchronization Signal}
\newacronym{tb}{TB}{Transport Block}
\newacronym{tcp}{TCP}{Transmission Control Protocol}
\newacronym{tdd}{TDD}{Time Division Duplexing}
\newacronym{tdma}{TDMA}{Time Division Multiple Access}
\newacronym{tfl}{TfL}{Transport for London}
\newacronym{tm}{TM}{Transparent Mode}
\newacronym{prr}{PRR}{Packet Reception Ratio}
\newacronym{trp}{TRP}{Transmitter Receiver Pair}
\newacronym{tti}{TTI}{Transmission Time Interval}
\newacronym{ttt}{TTT}{Time-to-Trigger}
\newacronym{tx}{TX}{Transmitter}
\newacronym{ue}{UE}{User Equipment}
\newacronym{ul}{UL}{Uplink}
\newacronym{uml}{UML}{Unified Modeling Language}
\newacronym{um}{UM}{Unacknowledged Mode}
\newacronym{utc}{UTC}{Urban Traffic Control}
\newacronym{vm}{VM}{Virtual Machine}
\newacronym{rsrq}{RSRQ}{Reference Signal Received Quality}
\newacronym{rssi}{RSSI}{Received Signal Strength Indicator}
\newacronym{crs}{CRS}{Cell Reference Signal}
\newacronym{v2v}{V2V}{Vehicle-to-Vehicle}
\newacronym{v2i}{V2I}{Vehicle-to-Infrastructure}
\newacronym{v2n}{V2N}{Vehicle-to-Network}
\newacronym{v2x}{V2X}{Vehicle-to-Everything}
\newacronym{vn}{VN}{Vehicular Node}
\newacronym{dsrc}{DSRC}{Dedicated Short Range Communication}
\newacronym{ci}{CI}{context information}
\newacronym{voi}{VoI}{value of information}
\newacronym{gps}{GPS}{Global Positioning System}
\newacronym{qos}{QoS}{Quality of Service}
\newacronym{qoe}{QoE}{Quality of Experience}
\newacronym{ml}{ML}{Machine Learning}
\newacronym{ahp}{AHP}{Analytic Hierarchy Process}
\newacronym{lidar}{LiDAR}{Light Detection and Ranging}
\newacronym{radar}{RADAR}{Radio Detection and Ranging}
\newacronym{sumo}{SUMO}{Simulation of Urban MObility}
\newacronym{wave}{WAVE}{Wireless Access in Vehicular Environment}
\newacronym{c-its}{C-ITS}{Connected Intelligent Transportation System}
\newacronym{dash}{DASH}{Dynamic Adaptive Streaming over HTTP}
\newacronym{http}{HTTP}{HyperText Transfer Protocol}
\newacronym{3d}{3D}{three-dimensional}
\newacronym{2d}{2D}{bi-dimensional}
\newacronym{pcc}{PCC}{Point Cloud Compression}
\newacronym{vg}{VG}{Voxel Grid}
\tikzstyle{startstop} = [rectangle, rounded corners, minimum width=2cm, minimum height=0.5cm,text centered, draw=black]
\tikzstyle{io} = [trapezium, trapezium left angle=70, trapezium right angle=110, minimum width=3cm, minimum height=1cm, text centered, draw=black]
\tikzstyle{process} = [rectangle, minimum width=2cm, minimum height=0.5cm, text centered, draw=black, alignb=center]
\tikzstyle{decision} = [ellipse, minimum width=2cm, minimum height=1cm, text centered, draw=black]
\tikzstyle{arrow} = [thick,<->,>=stealth]
\tikzstyle{line} = [thick,>=stealth]
\tikzstyle{darrow} = [thick,<->,>=stealth,dashed]
\tikzstyle{sarrow} = [thick,->,>=stealth]
\tikzstyle{larrow} = [line width=0.1mm,dashdotted,->,>=stealth]

\makeatletter
\def\grd@save@target#1{%
  \def\grd@target{#1}}
\def\grd@save@start#1{%
  \def\grd@start{#1}}
\tikzset{
  grid with coordinates/.style={
    to path={%
      \pgfextra{%
        \edef\grd@@target{(\tikztotarget)}%
        \tikz@scan@one@point\grd@save@target\grd@@target\relax
        \edef\grd@@start{(\tikztostart)}%
        \tikz@scan@one@point\grd@save@start\grd@@start\relax
        \draw[minor help lines] (\tikztostart) grid (\tikztotarget);
        \draw[major help lines] (\tikztostart) grid (\tikztotarget);
        \grd@start
        \pgfmathsetmacro{\grd@xa}{\the\pgf@x/1cm}
        \pgfmathsetmacro{\grd@ya}{\the\pgf@y/1cm}
        \grd@target
        \pgfmathsetmacro{\grd@xb}{\the\pgf@x/1cm}
        \pgfmathsetmacro{\grd@yb}{\the\pgf@y/1cm}
        \pgfmathsetmacro{\grd@xc}{\grd@xa + \pgfkeysvalueof{/tikz/grid with coordinates/major step x}}
        \pgfmathsetmacro{\grd@yc}{\grd@ya + \pgfkeysvalueof{/tikz/grid with coordinates/major step y}}
        \foreach \x in {\grd@xa,\grd@xc,...,\grd@xb}
        \node[anchor=north] at (\x,\grd@ya) {\pgfmathprintnumber{\x}};
        \foreach \y in {\grd@ya,\grd@yc,...,\grd@yb}
        \node[anchor=east] at (\grd@xa,\y) {\pgfmathprintnumber{\y}};
      }
    }
  },
  minor help lines/.style={
    help lines,
    gray,
    line cap =round,
    xstep=\pgfkeysvalueof{/tikz/grid with coordinates/minor step x},
    ystep=\pgfkeysvalueof{/tikz/grid with coordinates/minor step y}
  },
  major help lines/.style={
    help lines,
    line cap =round,
    line width=\pgfkeysvalueof{/tikz/grid with coordinates/major line width},
    xstep=\pgfkeysvalueof{/tikz/grid with coordinates/major step x},
    ystep=\pgfkeysvalueof{/tikz/grid with coordinates/major step y}
  },
  grid with coordinates/.cd,
  minor step x/.initial=.5,
  minor step y/.initial=.2,
  major step x/.initial=1,
  major step y/.initial=1,
  major line width/.initial=1pt,
}
\makeatother
\usetikzlibrary{shapes,arrows,chains}

\tikzstyle{flowchartbase} = [draw, align=center, on chain]
\tikzstyle{decision} = [diamond, flowchartbase]
\tikzstyle{block} = [rectangle, flowchartbase]
\tikzstyle{io} = [trapezium, trapezium left angle=70,trapezium right angle=-70, flowchartbase]

\IEEEoverridecommandlockouts
\newcommand\copyrightnotice{%
\begin{tikzpicture}[remember picture,overlay]
\node[anchor=south,yshift=10pt] at (current page.south) {\fbox{\parbox{\dimexpr\textwidth-\fboxsep-\fboxrule\relax}{
\footnotesize \textcopyright 2022 IEEE. Personal use of this material is permitted.
Permission from IEEE must be obtained for all other uses, in any current or future media,
including reprinting/republishing this material for advertising or promotional purposes,
creating new collective works, for resale or redistribution to servers or lists,
or reuse of any copyrighted component of this work in other works.}}};
\end{tikzpicture}
}

\usepackage[capitalize]{cleveref}

\makeglossaries
\begin{document}

\title{\vspace{0.1cm}Point Cloud Compression for Efficient \\ Data Broadcasting: A Performance Comparison}

\author{\IEEEauthorblockN{Francesco Nardo, Davide Peressoni, Paolo Testolina, Marco Giordani, Andrea Zanella}
        \IEEEauthorblockA{Department of Information Engineering, University of Padova, Italy, email: \texttt{\{name.surname\}@dei.unipd.it}\\
        }
}

\maketitle

\copyrightnotice

\glsunset{nr}

\begin{abstract}
The worldwide commercialization of \gls{5G} wireless networks and the exciting possibilities offered by \glspl{cav} are pushing toward the deployment of heterogeneous sensors for tracking dynamic objects in the automotive environment.
Among them, \gls{lidar} sensors are witnessing a surge in popularity as their application to vehicular networks seem particularly promising.
\glspl{lidar} can indeed produce a \gls{3d} mapping of the surrounding environment, which can be used for object detection, recognition, and topography.
These data are encoded as a point cloud which, when transmitted, may pose significant challenges to the communication systems as it can easily congest the wireless channel.
Along these lines, this paper investigates how to compress point clouds in a fast and efficient way.
Both 2D- and a 3D-oriented approaches are considered, and the performance of the corresponding techniques is analyzed in terms of (de)compression time, efficiency, and quality of the decompressed frame compared to the original.
We demonstrate that, thanks to the matrix form in which \gls{lidar} frames are saved,  compression methods that are typically applied for 2D images give equivalent results, if not better, than those specifically designed for 3D point clouds.

\end{abstract}

\begin{tikzpicture}[remember picture,overlay]
\node[anchor=north,yshift=-10pt] at (current page.north) {\parbox{\dimexpr\textwidth-\fboxsep-\fboxrule\relax}{
\centering\footnotesize This paper has been accepted for presentation at IEEE Wireless Communications and Networking Conference (WCNC) 2022. \textcopyright 2022 IEEE.\\
Please cite it as: F. Nardo, D. Peressoni, P. Testolina, M. Giordani, and A. Zanella, “Point cloud compression for efficient data broadcasting: A performance comparison,” IEEE Wireless Communications and Networking Conference (WCNC), Austin, USA, 2022.}};
\end{tikzpicture}

\section{Introduction}
\label{sec:introduction}
\glsresetall
The \gls{lidar} sensor is a remote scanner that determines the distance with an object by measuring the time between the emission of a light pulse and the reception of the back-scattered signal.
\gls{LiDAR} pulses are generated by an array of lasers that fire thousands of times per second at different vertical inclinations, and that continuously rotate to produce  a  \gls{3d} omnidirectional representation of the surrounding environment in the form of a \emph{point cloud}.
Specifically, a \gls{lidar} point cloud consists of a set of \gls{3d} data points in space corresponding to the projections of the laser beams on the surface of shapes or objects, and may also provide
additional information including the laser intensity, scan angle, and reflectance properties of the surface.

In the last decades, \glspl{LiDAR} have been extensively applied to different research fields, including  agriculture (e.g., for topographic analysis and prediction of soil properties), military (e.g., for ground surveillance, navigation, search and rescue) and architecture (e.g., for detecting subtle topographic features).
More recently, LiDAR scanners have also been playing an increasingly important role for \glspl{cav} to enhance detection and recognition of road entities, and enable a safer driving environment~\cite{lu2014connected}.
Compared to other types of sensors such as RADARs or color/thermal cameras~\cite{yue2018lidar}, \glspl{lidar} are robust under almost all lighting and weather conditions, with or without glare and shadows, and are currently the most precise sensors to measure range~\cite{li2020lidar}.
On the other hand, LiDAR acquisitions may produce very large volumes of data that can be challenging to handle with standard \gls{v2x} technologies~\cite{giordani2018feasibility,giordani2019lte}. One possible method to solve capacity issues is by leveraging the \gls{mmwave} spectrum, as promoted by recent IEEE and 3GPP standardization activities for future vehicular networks~\cite{zugno2020toward}.
At the same time, sensor data should be carefully selected as a function of the available channel bandwidth, so as to save (already limited) network resources for the most valuable transmissions~\cite{giordani2019framework}. However, this typically requires machine learning methods to be trained and validated for identifying the critical data, which may be difficult to do on board of vehicles~\cite{giordani2019investigating}.
In any case, data rates could be further reduced if  the LiDAR point clouds were efficiently compressed before data are validated and broadcast~\cite{varischio2021hybrid}.

  \begin{figure*}[t!]
  \setlength{\belowcaptionskip}{-0.33cm}
    \centering
    \includegraphics[width=0.85\textwidth]{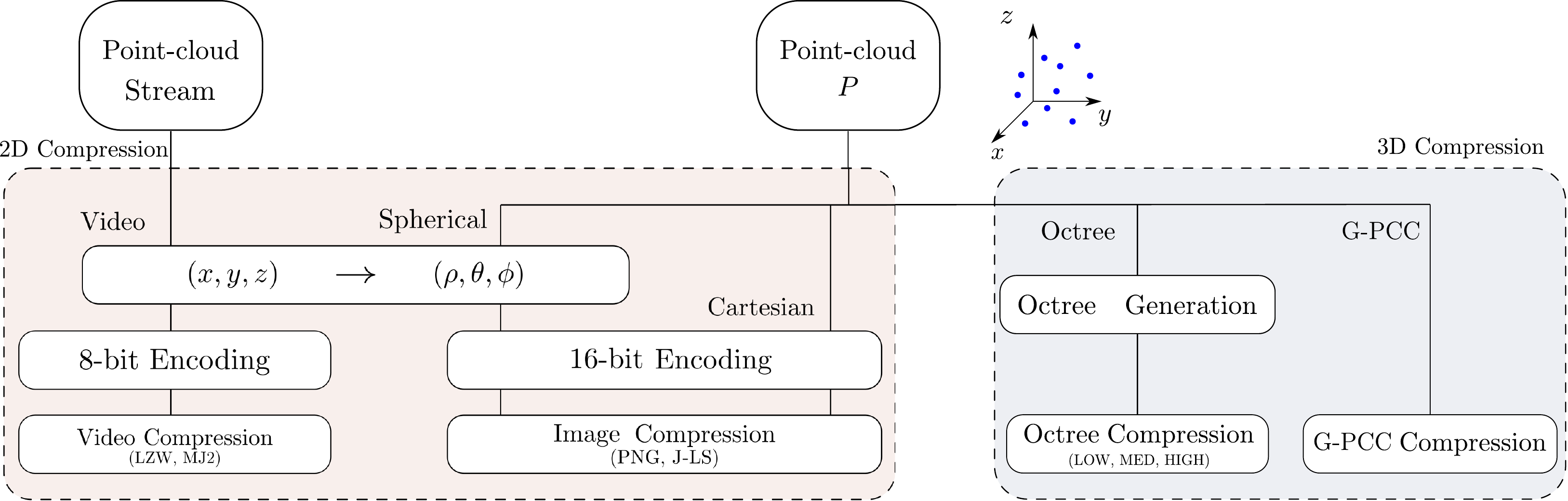}
    \caption{A schematic representation of the 3D and 2D compression methods reviewed in this paper.}
    \label{fig:scheme}
  \end{figure*}

In these regards, the most challenging aspect for data compression lies in the way the point cloud is represented. 
Given the \gls{3d} nature of LiDAR perceptions, geometric compression algorithms, based on \gls{pcd}, LASComp/LASzip~\cite{LASzip} and Octree~\cite{octree} formats, are the most common in the literature. 
More recent techniques based on deep learning, e.g., OctSqueeze~\cite{huang2020octsqueeze}, have been developed to enhance compressibility in 3D scenes.
Even though these methods preserve accuracy after compression, they require point-level processing of data, which may not be implemented in real time.
As a result, the scientific community is considering applying \gls{2d} transformations to the point cloud, using graph algorithms, and then exploit image-oriented compression techniques, such as \gls{J-LS}~\cite{loco-i} or \gls{PNG}~\cite{PNG}, as well as video-oriented and dictionary-based compression techniques, such as \gls{MJ2}~\cite{MJ2-ITU} and \gls{LZW}~\cite{LZW} respectively, to reduce computational complexity.
Despite~these studies, however, there is no accepted standard for point cloud compression, thus stimulating further~research. 

Based on the above introduction, in this paper we provide a comparison between 2D and 3D compression methods for point clouds, shedding light on the most promising scheme(s) to guarantee accurate though efficient compression before data broadcasting.
Compared to prior work, e.g.,~\cite{img}, our performance analysis is assessed not only in terms of average compression ratio (which generally indicates how accurately data is compressed), but also in terms of \gls{PSNR} (which measures the quality of the compressed point cloud, a critical requirement to ensure precise autonomous driving operations) and (de)compression time (to verify whether the point cloud can be (de)compressed in low latency, as is the case in safety-critical applications). Moreover, we study both 3D and image/video-based compression strategies, and investigate whether representing the point cloud with spherical coordinates, as natively supported in LiDAR data, would result in better compression than considering Cartesian coordinates.
Our simulation results, validated on the public Veloview Sample Dataset, demonstrate that 2D compression methods are orders of magnitude more efficient and up to $20\times$ faster than the considered 3D methods, especially when spherical coordinates are adopted, while incurring limited accuracy degradation in the reconstructed point cloud.

The remainder of this paper is organized as follows. In Sec.~\ref{sec:3d} and  Sec.~\ref{sec:2d} we describe some of the most common 2D and 3D methods, respectively, to compress LiDAR point clouds. In Sec.~\ref{sec:performance_comparison} we present our simulation setup and numerical results. Finally, in Sec.~\ref{sec:conclusions_and_future_work} we summarize our main conclusions and suggestions for future work.



\section{3D Compression Methods} 
\label{sec:3d}
\glspl{LiDAR} emit light pulses and record the backscattered waveforms.
In general, from each return pulse, we can estimate the Cartesian coordinates $(x,y,z)$ and the angle of arrival of each point, the received signal intensity, the registered time, as well as other side information.
In this work, we consider the data returned by the Velodyne sensors, i.e., a collection of \gls{udp} packets encoded in a \gls{PCAP} file, and try to compress the file size.  

\subsection{3D Data Representation} 

The most challenging aspect of collecting point clouds is related to their unordered and sparse structure, which makes classical storage methods inefficient.
For this reason, new solutions have been specifically designed to represent \gls{point-cloud} data.
\Glspl{octree}~\cite{trees}, in particular, are an extension of binary trees in which each internal node has exactly eight children, and that can be used to partition \gls{3d} spaces.
The root of the Octree is associated to the bounding box containing the whole point cloud.
Then, the space volume is partitioned in eight parts, each assigned to a children of the root node.
Each level of the space is subsequently split in eight parts.
Thus, each child represents $1/2^3$ of the parent space.
With this approach, each point collected by the \gls{LiDAR} is represented by the leaf which contains it, so the encoding precision grows with the number of levels.
A similar structure is the \gls{vg} that has been traditionally used in computer graphics to reduce both the input space dimensionality and the number of points in the raw point cloud.
The \gls{vg} sub-sampling technique is based on a grid of 3D \emph{\glspl{voxel}}.\footnote{A \gls{voxel} is a discrete volumetric element used in the visualization and analysis of 3D data. It represents the equivalent of a 2D-image pixel but on a regular grid in the 3D space.}
For each \gls{voxel}, a centroid is chosen as the representative of all the points that lie on the corresponding partition of the space.~\cite{VoxelLidar}.
Clearly, both the Octree and the VG representations introduce a quantization error, which depends on the granularity of the space partition.

\subsection{3D Data Compression} 
Several 3D compression algorithms were developed, depending on how the point cloud is represented.
For example, a compression algorithm exploiting the Octree data structure to perform predictive decoding based on local surface approximations was proposed in~\cite{octree}.
A deep neural network, also based on Octree data, was then introduced in~\cite{huang2020octsqueeze}.
Notably, a fast compression algorithm was developed in~\cite{VoxelSpheric} considering spherical \glspl{voxel}, while the Moving Picture Expert Group (MPEG) has released specifications for the  video-based (V-PCC) and the geometry-based (G-PCC) point cloud compression standards~\cite{graziosi2020overview}.

In this work, we analyze the efficiency and accurateness of G-PCC, as a possible standard for 3D point cloud compression, and of the Octree representation, as an efficient 3D storage method for point clouds, as illustrated in Fig.~\ref{fig:scheme}.
For the Octree generation, the \gls{PCL}\footnote{The \gls{PCL} is a standalone, large scale, open C++ library for point cloud processing and management. The PCL can be publicly accessed at \url{https://pointclouds.org/documentation/tutorials/pcd\_file\_format.html}.}~\cite{PCL} was used.
Each node of the Octree is represented by 8 bits, each stating whether the corresponding space partition is empty ($0$) or contains at least one point ($1$).
Then, all non-zero bytes are saved in breadth-first order \cite{VoxelLidar}.
\gls{PCL} offers 12 different resolution profiles (corresponding to 12 different levels of compression), which can be grouped into 3 categories, i.e., HIGH, MEDIUM, and LOW, as reported in Table~\ref{tab:profili_octree}.

  \begin{table}[t!]
  \centering
   \vspace{0.15cm}
  \includegraphics[width=0.9\columnwidth]{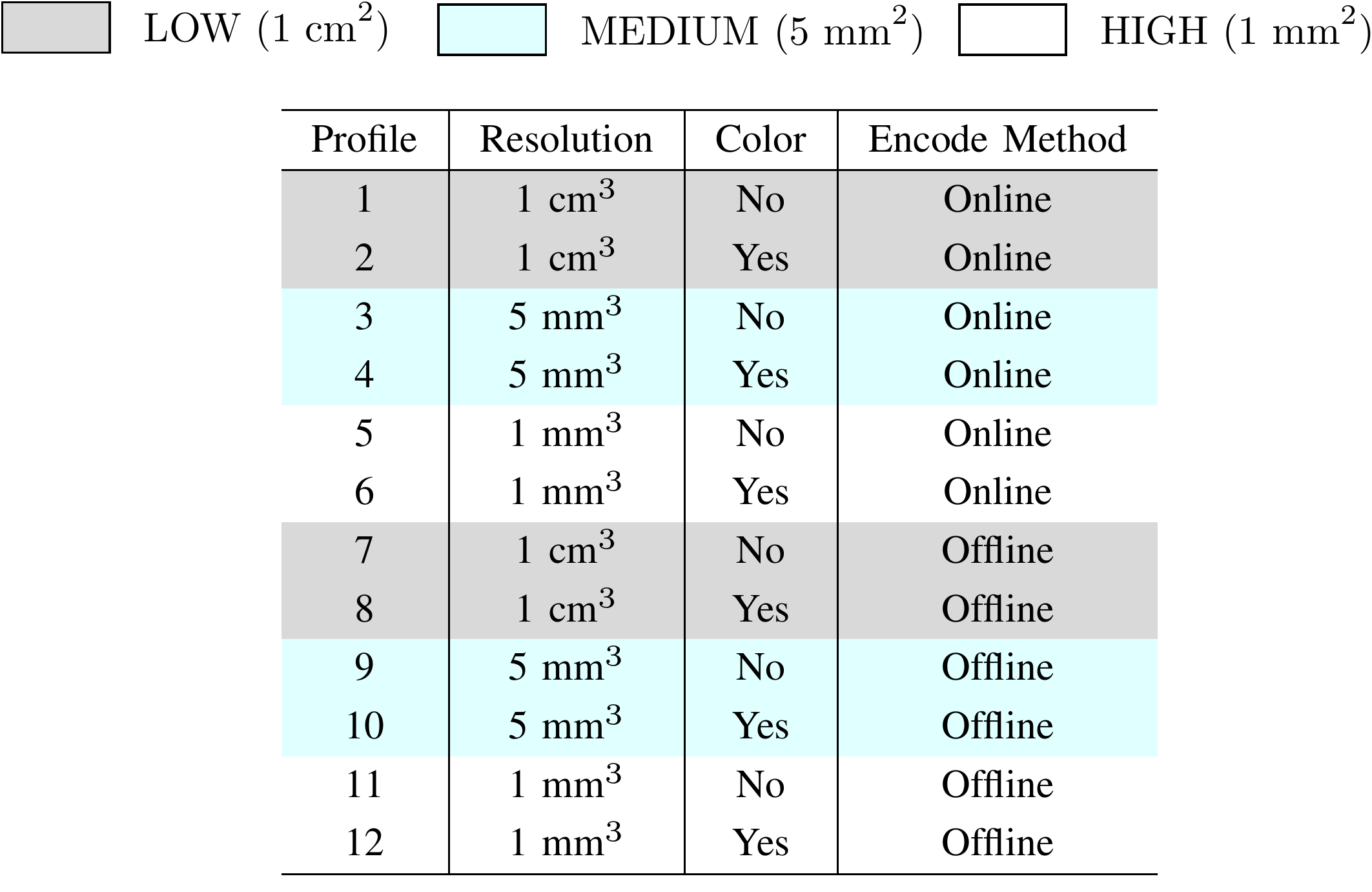}
  \captionof{table}{List of Octree compression profiles according to the \gls{PCL}~\cite{PCL}.}
   \label{tab:profili_octree}
    \renewcommand{\arraystretch}{1.3}
    \vspace{-0.33cm}
  \end{table}

\section{2D Compression Methods} 
\label{sec:2d}

In this section we discuss how 3D LiDAR point clouds can be transformed into 2D representations (Sec.~\ref{ssec:more_images}), and then compressed via 2D methods originally designed to compress images (Sec.~\ref{ssec:2d_based_compression_image}) and videos (Sec.~\ref{ssec:video_based_compression}).

  \subsection{3D-to-2D Data Representation}\label{ssec:more_images}
  The 3D \gls{LiDAR} data can be stored into a 2D image array, represented through Cartesian or spherical coordinates. 
  \begin{itemize}
    \item \emph{Cartesian representation.} According to prior work~\cite{img}, the original Cartesian $(x,y,z)$ point-cloud coordinates can be mapped into the \gls{2d} plane according to one of the following strategies:
  \begin{itemize}
    \item \textit{Single-channel Cartesian representation}:
      The points are saved in three single-channel (grayscale) images by assigning Cartesian coordinates to each image.
     \item \textit{Tri-channel Cartesian representation}:
      The points are saved in one \gls{RGB} colored image by assigning the $x$ coordinate to the R channel, the $y$  coordinate to the G channel, and the $z$ coordinate to the B channel.
  \end{itemize}
  \item \emph{ Spherical representation.} We propose to represent the point cloud through the $(\rho,\theta,\phi)$ spherical coordinates (where $\rho$ is the radial distance, $\theta$ is the polar/elevation angle and $\phi$ is the azimuth angle), computed as $\rho = \sqrt{x^2 + y^2 + z^2}$, $\theta = \arctan{\left(z/(\sqrt{x^2+y^2})\right)}$, and ${\phi = \arctan{(y/x)}}$.
  The data is then stored into a 2D image array.
  \end{itemize}

  
  Notice that converting 3D data into a 2D image requires the point-cloud coordinates to be converted from \texttt{float} to \texttt{unsigned integers}, thus introducing a quantization error.
  Considering a $n$-bit encoding, a floating point value $u_f$ can be easily converted to \texttt{unsigned integer} $u_i$ as
  \begin{equation}
    u_i = \left\lfloor{  {\frac{u_f - \min(u_f)}{\max(u_f) - \min(u_f)} \cdot (2^{n}-1)} }\right\rceil.
    \label{eq:normalization}
  \end{equation}
  In our trials, we observed that encoding with more than $16$ bits would not bring any significant improvement in terms of accuracy, thus we set $n=16$ in our simulations.

  This image representation of \gls{LiDAR} frames makes the application of existing 2D compression algorithms quite straightforward.
  Furthermore, the image encoding preserves the value continuity of the scene, i.e., neighboring pixels have similar values, an important property when applying image compression algorithms: each row in the matrix-form representation of the point cloud contains the points having the same elevation angle, i.e., acquired by the same laser, whereas the columns scan the azimuth space, according to the laser rotation.
  In the following, we present the 2D compression algorithms that we considered in this work, as illustrated in Fig.~\ref{fig:scheme}.


  \subsection{2D (Image-Based) Data Compression} 
  \label{ssec:2d_based_compression_image}
  For image-based compression, we consider the well-known \gls{png} and \gls{J-LS} image formats.
  Specifically, \gls{png} uses \gls{Deflate}, a compression algorithm that combines \gls{LZW}~\cite{LZW} with the Huffman coding~\cite{Huffman}.
  Similarly to other dictionary coders, \gls{LZW} employs a sliding window to scan the data: whenever a new sequence of bytes is observed, the corresponding dictionary entry is created and all the subsequent occurrences of the same sequence are substituted with the corresponding dictionary index.
  The dictionary is then compressed with the Huffman coding.

  The J-LS algorithm~\cite{loco-i}, instead, predicts the value of each pixel in the image from the values of the neighboring pixels, thus leveraging the correlation among consecutive frames. This information is then modeled through a two-sided geometric distribution, and encoded using the Golomb coding, which is similar to the Huffman one.



  \subsection{2D (Video-Based) Data Compression}
  \label{ssec:video_based_compression}
Once the 3D LiDAR frames are converted into their 2D representation, video-based compression techniques (either inter- or intra-frame) can be applied.
  Specifically, we analyze the performance of an adaptation of \gls{LZW} (\gls{Deflate}) for videos, and the \gls{MJ2} algorithm.
  In both cases, 8 bit-encoding was applied, as typically considered in the most common video encoding standards.
  
 We easily extended \gls{LZW}~\cite{LZW} (specifically \gls{Deflate}) in order to be applied inter-frame compression, i.e., taking into account the temporal correlation among consecutive frames to improve compression rates.
  Namely, for a given sequence of $N$ \gls{lidar} frames, three $N$-long vectors are generated, one per coordinate -- either Cartesian $(x,y,z)$ or spherical $(\rho,\theta,\phi)$.
  \gls{Deflate} is then applied to each coordinate vector separately.

  \gls{MJ2} is another popular video coding scheme~\cite{MJ2-ITU}.
  In this case, the video frames are generated as a sequence of images, according to the representation strategies described in Sec.~\ref{ssec:more_images}.
  Then, each frame is independently encoded using JPEG 2000. Because of the intra-frame encoding, the MJ2 is more resilient to propagation of errors over time, more scalable, and better suited to networked and point-to-point environments than DEFLATE. 
  Also, it permits random access to individual~frames.


\section{Performance Comparison} 
\label{sec:performance_comparison}

In this section we first describe our simulation scenario and performance metrics (Sec.~\ref{sub:simulation_scenarios_and_parameters}), then we present our main performance results (Sec.~\ref{sub:numerical_results}).

  \subsection{Simulation Scenario and Parameters}
  \label{sub:simulation_scenarios_and_parameters}
  The performance of compression algorithms has been compared on the Veloview Sample Dataset\footnote{The Veloview Sample Dataset can be publicly accessed at \url{https://data.kitware.com/\#collection/5b7f46f98d777f06857cb206}.}, that contains data from seven heterogeneous road environments acquired with a Velodyne VLP-16 and a Velodyne HDL-32 LiDAR, so as to consider different point cloud resolutions. In particular, the former sensor uses 16 laser beams with an angular resolution of 2 degrees and 0.1 degrees on the elevation and azimuth dimensions, respectively, while the latter configures up to 32 laser beams at around twice the resolution. Furthermore, the data was acquired using two rotation frequencies, i.e., 600 and 1200 rpm, thus further increasing the data diversity and the robustness of the results.
  The datasets were converted from the original PCAP format into the CSV or Binary PCD formats with the Matlab \texttt{velodyneFileReader} module, to be then used in our custom Python code for performance evaluation.

  We compare the performance of the compression algorithms reviewed in Secs.~\ref{sec:3d} and \ref{sec:2d}. For 3D methods, we consider the Octree compression levels (HIGH, MEDIUM, and LOW), and G-PCC with default parameters. For 2D methods, we compare image-based (PNG and J-LS, considering both tri-channel Cartesian and spherical representations) and video-based (LZW and MJ2, considering spherical representation only) solutions.
  The following metrics have been used to evaluate the compression algorithms.
    \paragraph{Compression rate} Let $\mathcal{\bar{S}}$ be the size of the compressed point cloud, and $\mathcal{{S}_{\rm raw}}$ be the size of the raw point cloud, which is the {PCAP} file from the \gls{LiDAR} acquisition. The compression rate measures the reduction in size of the data representation produced by compression, and is given by 
      \begin{equation}
        \text{Compression rate} = 1 - ({\mathcal{\bar{S}}}/{\mathcal{{S}_{\rm raw}}}).
      \end{equation}

    \paragraph{\gls{Bpp}} Let $|\bar{P}|$ be the total number of points contained in the compressed point cloud of size $\mathcal{\bar{S}}$. The \gls{Bpp} is defined as the number of bytes used to compress each point in the original point cloud, and is quantified as 
      \begin{equation}
        \text{BPP} = {\mathcal{\bar{S}}}/{|\bar{P}|}.
      \end{equation}

    \paragraph{Point-to-plane \acrlong{PSNR} (PSNR)}
    The \gls{PSNR} is proportional to the quality of reconstructed point clouds/images/videos subject to compression, and is thus related to the accuracy of autonomous driving operations like object detection~\cite{li2021deep}.
     Let $\bm{p}\in P$ be one point in the original point cloud $P$, and $\bm{q}\in \hat{P}$ be its nearest neighbor in the reconstructed point cloud $\hat{P}$. The point-to-plane \gls{MSE} can be computed with respect to $P$ as
\begin{equation}
  \mathrm{MSE}_{P\to \hat{P}} = \dfrac{1}{|P|}\sum_{\forall \bm{p}\in P} \left( \left\langle \bm{p}-\bm{q}, \bm{n}_{\bm{q}} \right\rangle \right)^2,
\label{eq:mse}
\end{equation}
where $\bm{n}_{\bm{q}}$ is the surface tangent in $\bm{q}\in \hat{P}$, and $\left\langle \bm{p}-\bm{q}, \bm{n}_{\bm{q}} \right\rangle$ is the projection of vector $\bm{p}-\bm{q}$ on $\bm{n}_{\bm{q}}$. Accordingly, the point-to-plane \gls{PSNR} with respect to $P$ can be written as
\begin{equation}
  \mathrm{PSNR}_{P\to \hat{P}} = 10 \log_{10}\left(\frac{(\theta_P^*)^2}{\mathrm{MSE}_{P\to \hat{P}}}\right)
\label{eq:psnr-ref}
\end{equation}
where $\theta_P^*$ represents the peak value in the original point cloud $P$.
Generally, $\theta_P^*$ is selected according to the nearest neighbor distances $d_p$ for all points $\mathbf{p}$ in $P$, i.e., $\theta_P^* = \max_{\forall\bm{p} \in P}\left\{ d_p \right\}$.
Then, the \gls{PSNR} between $P$ and $\hat{P}$ is given by
\begin{equation}
  \mathrm{PSNR}_{P,\hat{P}} = \min\left\{{\mathrm{PSNR}_{P\to \hat{P}},\ \mathrm{PSNR}_{\hat{P}\to P}}\right\}.
\label{eq:psnr}
\end{equation}

\paragraph{Computation time} It refers to the time required to compress/decompress the point cloud (from when the raw LiDAR output is produced in the form of a PCAP file, until the compressed point cloud is generated, or vice versa) using one of the techniques presented in the paper. 
This quantity has been measured on a machine executing an Intel Core i5-4210U processor at 1.70 GHz, running {Linux} 5.4.74-1, {Python} 3.8.6, {g++} 10.2.0, using {\gls{PCL}} 1.10 for Octree compression and G-PCC 13. All the trails have been run single threaded.


  \subsection{Numerical Results} 
  \label{sub:numerical_results}

  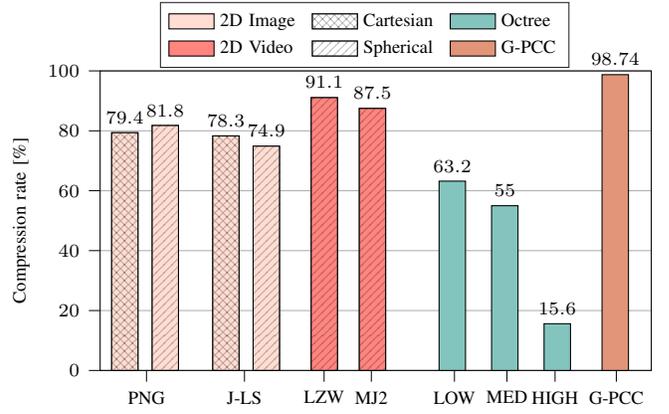
\begin{figure}[t!]
  \setlength{\belowcaptionskip}{-0.43cm}
    \centering
        \setlength\fwidth{0.87\columnwidth} 
    \setlength\fheight{0.45\columnwidth}
	\begin{tikzpicture}
	
\definecolor{color0}{RGB}{255,205,178}%
\definecolor{color1}{RGB}{255,180,162}%
\definecolor{color2}{RGB}{229,152,155}%
\definecolor{color3}{RGB}{181, 131, 141}%
\definecolor{color4}{RGB}{109, 104, 117}%

\definecolor{mycolor6}{RGB}{0, 109, 119}%
\definecolor{mycolor7}{RGB}{131, 197, 190}%
\definecolor{mycolor8}{RGB}{237, 246, 249}%
\definecolor{mycolor9}{RGB}{255, 221, 210}%
\definecolor{mycolor10}{RGB}{226, 149, 120}%
\definecolor{mycolor11}{RGB}{255, 255, 255}%
	\pgfplotsset{
tick label style={font=\scriptsize},
label style={font=\scriptsize},
legend  style={font=\scriptsize}
}

\begin{axis}[%
width=0.951\fwidth,
height=\fheight,
at={(0\fwidth,0\fheight)},
scale only axis,
legend cell align={center},
legend style={legend cell align=left, align=left, draw=white!15!black, at={(0.48,1.23)},/tikz/every even column/.append style={column sep=0.15cm},
  anchor=north ,legend columns=3},
tick align=outside,
tick pos=left,
ybar,
bar width=10pt,
x grid style={white!69.0196078431373!black},
xmin=0, xmax=6.3,
xtick style={color=black},
xtick={0.53,1.65,2.56,3.1,4.05,4.65,5.2,5.9},
xticklabels={PNG, J-LS, LZW, MJ2, LOW, MED, HIGH, G-PCC},
y grid style={white!69.0196078431373!black},
ylabel={Compression rate [\%]},
ymajorgrids,
ymin=0, ymax=100,
nodes near coords,
every node near coord/.append style={font=\scriptsize},  
nodes near coords align={vertical},  
ytick style={color=black},
x tick label style={rotate=0},
			]
			

			\addlegendimage{ybar,area legend,fill=mycolor9, draw=black};
			\addlegendentry{2D Image}

			\addlegendimage{ybar,area legend,fill=mycolor11, draw=black, postaction={pattern=crosshatch, opacity=0.5}};
			\addlegendentry{Cartesian}
		
			\addlegendimage{ybar,area legend,fill=mycolor7, draw=black};
			\addlegendentry{Octree}

			\addlegendimage{ybar,area legend,fill=mycolor9!60!red, draw=black};
			\addlegendentry{2D Video}

			\addlegendimage{ybar,area legend,fill=mycolor11, draw=black, postaction={pattern=north east lines, opacity=0.5}};
			\addlegendentry{Spherical}

			\addlegendimage{ybar,area legend,fill=mycolor10, draw=black};
			\addlegendentry{G-PCC}



			\addplot[draw=color0,fill=mycolor9, draw=black, postaction={pattern=crosshatch, opacity=0.5}] coordinates {
			(1.37,79.4)
			};
			\addplot[draw=color0,fill=mycolor9, draw=black, postaction={pattern=north east lines, opacity=0.5}] coordinates {
			(1.47,81.8) 
			};

			\addplot[draw=color0,fill=mycolor9, draw=black, postaction={pattern=crosshatch, opacity=0.5}] coordinates {
			(1.8,78.3)
			};
			\addplot[draw=color0,fill=mycolor9, draw=black, postaction={pattern=north east lines, opacity=0.5}] coordinates {
			(1.9,74.9)
			};

			\addplot[draw=color0,fill=mycolor9!60!red, draw=black,  postaction={pattern=north east lines, opacity=0.5}] coordinates {
			(2.2,91.1) (2.75,87.5) 
			};

			\addplot[draw=color0,fill=mycolor7, draw=black] coordinates {
			(3.3,63.2) (3.9,55) (4.5,15.6)
			};

			\addplot[draw=color0,fill=mycolor10, draw=black] coordinates {
			(4.8,98.74)
			};



		\end{axis}





\end{tikzpicture}
    \caption{Compression rate for different 2D vs. 3D compression methods.}
    \label{fig:savings-avg}
\end{figure}



\textbf{Compression efficiency.} In Fig.~\ref{fig:savings-avg} we plot the compression rate for different compression methods. 
First, we observe that G-PCC achieves the best compression rate ($98.74\%$), thus imposing as the standard for point cloud compression.
Second, 2D compression and, in particular, PNG and J-LS, outperforms the Octree-based compression. 
In fact, unlike their 2D counterparts, Octree methods tend to overfit the data and cannot detect and appropriately remove redundant information hidden in the point cloud representations~\cite{varischio2021hybrid}, resulting in a dramatic drop in the compression rate when increasing the resolution.
 On the contrary, PNG still guarantees a promising $80\%$ compression rate, up to $25\%$ better than Octree. 

Third, Fig.~\ref{fig:savings-avg} shows that representing the point cloud with spherical coordinates can result in better compression than using Cartesian coordinates, e.g., in case of PNG. 
In fact, while the former approach tries to store only radius and azimuth for each point in the LiDAR data (the elevation angle is indeed constant for each LiDAR laser beam), raw Cartesian files encode three geometric coordinates as a tri-channel image, thus using about 1/3 more BPPs than in the spherical methods. We also tried to convert the point cloud into spherical coordinates using the radius only, thus representing the LiDAR's input as a single-channel image. While this approach permits to reduce the BPPs by 2/3 compared to Cartesian files, the final compression rate was unsatisfactory.

Third, Fig.~\ref{fig:savings-avg} illustrates that video-based methods like LZW can compress efficiently by taking advantage of the temporal correlation between neighboring frames in the 2D point cloud representation, for example tracking the movement of cars: compared to PNG, LZW achieves a $+11\%$ improvement, just $8\%$ less than G-PCC.

  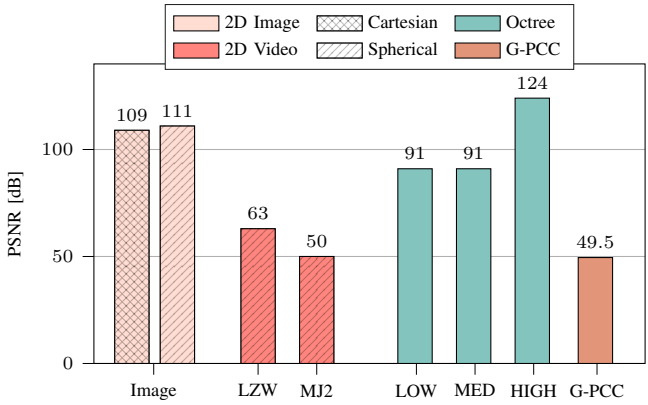
\begin{figure}[t!]
  \setlength{\belowcaptionskip}{-0.43cm}
    \centering
        \setlength\fwidth{0.87\columnwidth} 
    \setlength\fheight{0.45\columnwidth}
	\begin{tikzpicture}
	
\definecolor{color0}{RGB}{255,205,178}%
\definecolor{color1}{RGB}{255,180,162}%
\definecolor{color2}{RGB}{229,152,155}%
\definecolor{color3}{RGB}{181, 131, 141}%
\definecolor{color4}{RGB}{109, 104, 117}%

\definecolor{mycolor6}{RGB}{0, 109, 119}%
\definecolor{mycolor7}{RGB}{131, 197, 190}%
\definecolor{mycolor8}{RGB}{237, 246, 249}%
\definecolor{mycolor9}{RGB}{255, 221, 210}%
\definecolor{mycolor10}{RGB}{226, 149, 120}%
	\pgfplotsset{
tick label style={font=\scriptsize},
label style={font=\scriptsize},
legend  style={font=\scriptsize}
}

\begin{axis}[%
width=0.951\fwidth,
height=\fheight,
at={(0\fwidth,0\fheight)},
scale only axis,
legend cell align={left},
legend style={legend cell align=left, align=left, draw=white!15!black, at={(0.5,1.2)},/tikz/every even column/.append style={column sep=0.15cm},
  anchor=north ,legend columns=3},
tick align=outside,
tick pos=left,
ybar,
bar width=13pt,
x grid style={white!69.0196078431373!black},
xmin=0, xmax=4.7,
xtick style={color=black},
xtick={0.51,1.4,1.9,2.75,3.25,3.75,4.29},
xticklabels={Image, LZW, MJ2, LOW, MED, HIGH, G-PCC},
y grid style={white!69.0196078431373!black},
ylabel={PSNR [dB]},
ymajorgrids,
ymin=0, ymax=140,
nodes near coords,
every node near coord/.append style={font=\scriptsize},  
nodes near coords align={vertical},  
ytick style={color=black},
x tick label style={rotate=0},
			]
			

			\addlegendimage{ybar,area legend,fill=mycolor9, draw=black};
			\addlegendentry{2D Image}

			\addlegendimage{ybar,area legend,fill=white, draw=black, postaction={pattern=crosshatch, opacity=0.5}};
			\addlegendentry{Cartesian}
		
			\addlegendimage{ybar,area legend,fill=mycolor7, draw=black};
			\addlegendentry{Octree}

			\addlegendimage{ybar,area legend,fill=mycolor9!60!red, draw=black};
			\addlegendentry{2D Video}

			\addlegendimage{ybar,area legend,fill=white, draw=black, postaction={pattern=north east lines, opacity=0.5}};
			\addlegendentry{Spherical}

			\addlegendimage{ybar,area legend,fill=mycolor10, draw=black};
			\addlegendentry{G-PCC}



			\addplot[draw=color0,fill=mycolor9, draw=black,postaction={pattern=crosshatch, opacity=0.5}] coordinates {
			(1,109) 
			};

			\addplot[draw=color0,fill=mycolor9, draw=black,postaction={pattern=north east lines, opacity=0.5}] coordinates {
			(1.05,111) 
			};

			\addplot[draw=color0,fill=mycolor9!60!red, draw=black,  postaction={pattern=north east lines, opacity=0.5}] coordinates {
			(1.4,63) (1.9,50) 
			};

			\addplot[draw=color0,fill=mycolor7, draw=black] coordinates {
			(2.4,91) (2.9,91) (3.4,124)
			};

			\addplot[draw=color0,fill=mycolor10, draw=black] coordinates {
			(3.6,49.5)
			};



		\end{axis}





\end{tikzpicture}
    \caption{PSNR for different 2D vs. 3D compression methods. ``Image'' compression is obtained by averaging PNG and J-LS compression.}
    \label{fig:psnr}
\end{figure}



\textbf{Compression accuracy.}
Compression accuracy is measured in terms of \gls{PSNR}, as depicted in Fig.~\ref{fig:psnr} (where the ``Image'' bars are obtained by averaging PNG and J-LS schemes, that gave similar results). 
It appears clear that Octree with HIGH profile exhibits the best performance ($+14\%$ against PNG, however in the face of a significant degradation in terms of compression rate), even though both LOW and MEDIUM profiles underperform image-based methods ($-17\%$).
In any case, the PSNR is guaranteed to be above 100 dB, thereby resulting in basically \emph{lossless} compression; this ensures that the reconstructed point cloud after decompression can be considered the same as the original dataset.
Notably, for image-based methods, both Cartesian and spherical representations give similar PSNR performance.

On the other hand, Fig.~\ref{fig:psnr} shows that video-based compression, despite the high compression rate, suffers from very bad accuracy compared to both image- (up to $-55\%$) and Octre-based (up to $-60\%$) schemes. In fact, while static images are encoded with 16 bits, video frames are designed to operated with 8 bits, as illustrated in Sec.~\ref{ssec:video_based_compression}.
Even though updates to both LZW and MJ2 standards have been made to increase the bit-depth, commercially available implementations are still limited to 8 (or sometimes 12) bits per sample, which make the compression \emph{lossy}.

Similarly, G-PCC exhibits a low \gls{PSNR}, thus revealing the accuracy cost (up to 74 dB vs. Octree and 60 dB vs. 2D solutions) required to achieve its outstanding compression rate.

  \begin{figure}[t!]
   \centering
   \setlength{\belowcaptionskip}{-0.7cm}
  \begin{subfigure}[t!]{0.44\textwidth}
  \centering
  \setlength{\belowcaptionskip}{-0.42cm}
   \setlength\fwidth{0.9\columnwidth} 
    \setlength\fheight{0.4\columnwidth}
    \input{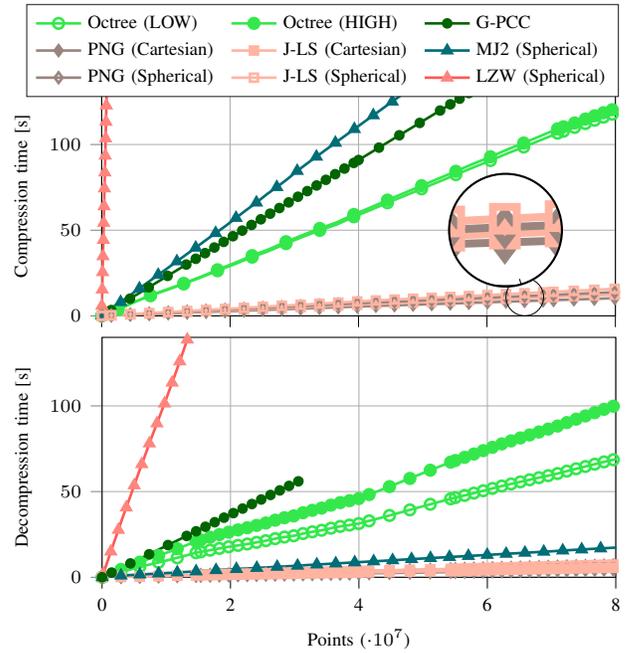}
     \label{fig:times-c}
     \vspace{-1.35\baselineskip}
   \end{subfigure}
   \begin{subfigure}[t!]{0.44\textwidth}
  \centering
  \setlength{\belowcaptionskip}{-3cm}
   \setlength\fwidth{0.9\columnwidth} 
    \setlength\fheight{0.4\columnwidth}
    \pgfplotsset{scaled x ticks=false}
  \begin{tikzpicture}[spy using outlines=
{rounded rectangle, magnification=4, connect spies}]

\definecolor{color0}{RGB}{255,205,178}%
\definecolor{color1}{RGB}{255,180,162}%
\definecolor{color2}{RGB}{229,152,155}%
\definecolor{color3}{RGB}{181, 131, 141}%
\definecolor{color4}{RGB}{109, 104, 117}%

\definecolor{mycolor6}{RGB}{0, 109, 119}%
\definecolor{mycolor7}{RGB}{131, 197, 190}%
\definecolor{mycolor8}{RGB}{237, 246, 249}%
\definecolor{mycolor9}{RGB}{255, 221, 210}%
\definecolor{mycolor10}{RGB}{226, 149, 120}%
\definecolor{mycolor11}{RGB}{0, 100, 0}%
  \pgfplotsset{
tick label style={font=\scriptsize},
label style={font=\scriptsize},
legend  style={font=\scriptsize},
filter discard warning=false
}

\begin{axis}[%
width=0.951\fwidth,
height=\fheight,
at={(0\fwidth,0\fheight)},
scale only axis,
legend cell align={left},
tick align=outside,
tick pos=left,
x grid style={white!69.0196078431373!black},
xmajorgrids,
xmin=0, xmax=80000000,
xtick={0,20000000,40000000,60000000,80000000},
xticklabels={0, 2, 4, 6,8},
xtick style={color=black},
ymin = 0, ymax=140,
xlabel={Points ($\cdot10^7$)},
y grid style={white!69.0196078431373!black},
ylabel={Decompression time [s]},
ymajorgrids,
      ]

\addplot [color=mycolor7!40!green, line width=1.0pt, mark size=2pt, mark=o, mark options={solid, fill=mycolor7!40!green, mycolor7!40!green},mark repeat=2]
  table[x= Points, y= LOW D, col sep=comma]{grafici/times.csv};

\addplot [color=mycolor7!40!green, line width=1.0pt, mark size=2pt, mark=*, mark options={solid, fill=mycolor7!40!green, mycolor7!40!green},mark repeat=2]
  table[x= Points, y= HIGH D, col sep=comma]{grafici/times.csv};

\addplot [color=mycolor9!60!black, line width=1.0pt, mark size=2pt, mark=diamond*, mark options={solid, fill=mycolor9!60!black, mycolor9!60!black},mark repeat=2]
  table[x= Points, y= PNG car D, col sep=comma]{grafici/times.csv};

\addplot [color=mycolor9!60!black, line width=1.0pt,  mark size=2pt, mark=diamond, mark options={solid, fill=mycolor9!60!black, mycolor9!60!black},mark repeat=2]
  table[x= Points, y= PNG sph D, col sep=comma]{grafici/times.csv};

\addplot [color=color1, line width=1.0pt, mark size=1.5pt, mark=square*, mark options={solid, fill=color1, color1},mark repeat=2]
  table[x= Points, y= J-LS car D, col sep=comma]{grafici/times.csv};

\addplot [color=color1, line width=1.0pt,  mark size=1.5pt, mark=square, mark options={solid, fill=color1, color1},mark repeat=2]
  table[x= Points, y= J-LS sph D, col sep=comma]{grafici/times.csv};

\addplot [color=mycolor6, line width=1.0pt,  mark size=2pt, mark=triangle*, mark options={solid, fill=mycolor6, mycolor6},mark repeat=2]
  table[x= Points, y= Video D, col sep=comma]{grafici/times.csv};

\addplot [color=mycolor9!40!red, line width=1.0pt,  mark size=2pt, mark=triangle*, mark options={solid, fill=mycolor9!60!red, mycolor9!60!red},mark repeat=2]
  table[x= Points, y= LZW D, col sep=comma,each nth point=20]{grafici/times.csv};

\addplot [color=mycolor11, line width=1.0pt,  mark size=1.5pt, mark=*, mark options={solid, fill=mycolor11, mycolor11},mark repeat=2]
  table[row sep=crcr]{%
0 0 \\
29216 0.053499048 \\
1489728 2.727924065 \\
2950944 5.403638215 \\
4412224 8.07946956  \\
5873664 10.75559389 \\
7335680 13.43277296 \\
8799712 16.11364365 \\
10262464  18.79217046 \\
11707104  21.43753137 \\
13172704  24.1212733  \\
14635200  26.79933133 \\
14649680  26.82584647 \\
15372624  28.14966957 \\
16096816  29.47577795 \\
16820832  30.80156404 \\
17544256  32.1262661  \\
18269104  33.45357571 \\
18992304  34.77786759 \\
19715040  36.1013098  \\
20438992  37.42697871 \\
21162608  38.75203234 \\
21886592  40.07775984 \\
22610592  41.40351664 \\
23334560  42.72921484 \\
24058096  44.05412198 \\
24782064  45.37982018 \\
25506800  46.70692471 \\
26230128  48.03145097 \\
26954624  49.35811602 \\
27677568  50.68193912 \\
28401680  52.007901 \\
29124608  53.3316948  \\
29848304  54.65689493 \\
30571088  55.98042504 \\
31297120  57.30990275 \\
};


\coordinate (spypoint) at (axis cs:65887200,6);
\coordinate (magnifyglass) at (axis cs:62887200,120);

    \end{axis}


  \end{tikzpicture}
     \label{fig:times-d}
  \end{subfigure}\hspace{-0.01cm}%
  \vspace{-0.43cm}
  \caption{Compression (above) and decompression (below) times for different 2D vs. 3D compression methods.}
     \label{fig:times}
     \end{figure}

\textbf{(De)compression time.}
Timely compression and decompression is of utmost importance for communication systems to ensure that sensor data is broadcast in real time.
From Fig.~\ref{fig:times} (above), we observe that image-based methods achieve up to $10\times$ and $20\times$ faster compression than Octree and G-PCC. In particular, PNG works slightly better than J-LS, achieving an improvement of $20\%$.
In both cases, the compression time grows linearly with the number of points in the point cloud, as expected.
On average, Octree and G-PCC can~compress around 670k and 440k point/s respectively, against the 5.5M  points/s for PNG. In comparison, the HDL-32 sensor captures 695k points/s, thereby making image-based compressors the only methods capable of processing the data at the frame rate of the LiDAR, thus achieving real-time performance.
Interestingly, video-based strategies (LZW and MJ2) are significantly slower than their competitors, which make them undesirable for most~applications.

In terms of decompression, Fig.~\ref{fig:times} (below) illustrates that image-based methods are still faster than the 3D ones. Notably, decompression takes less time than compression, a critical feature for autonomous driving since decompression is generally executed on-board of cars~\cite{varischio2021hybrid}.


  \begin{figure}[t!]
  \setlength{\belowcaptionskip}{-0.45cm}
    \centering
      \setlength\fwidth{0.87\columnwidth} 
    \setlength\fheight{0.6\columnwidth}
	\begin{tikzpicture}[spy using outlines=
{rounded rectangle, magnification=4, connect spies}]

\definecolor{color0}{RGB}{255,205,178}%
\definecolor{color1}{RGB}{255,180,162}%
\definecolor{color2}{RGB}{229,152,155}%
\definecolor{color3}{RGB}{181, 131, 141}%
\definecolor{color4}{RGB}{109, 104, 117}%

\definecolor{mycolor6}{RGB}{0, 109, 119}%
\definecolor{mycolor7}{RGB}{131, 197, 190}%
\definecolor{mycolor8}{RGB}{237, 246, 249}%
\definecolor{mycolor9}{RGB}{255, 221, 210}%
\definecolor{mycolor10}{RGB}{226, 149, 120}%
\definecolor{mycolor11}{RGB}{0, 100, 0}%
	\pgfplotsset{
tick label style={font=\scriptsize},
label style={font=\scriptsize},
legend  style={font=\scriptsize}
}

\begin{axis}[%
width=0.951\fwidth,
height=\fheight,
at={(0\fwidth,0\fheight)},
scale only axis,
legend cell align={left},
legend style={legend cell align=left, align=left, draw=white!15!black, at={(0.8,0.65)},/tikz/every even column/.append style={column sep=0.15cm},
  anchor=north ,legend columns=1},
tick align=outside,
tick pos=left,
x grid style={white!69.0196078431373!black},
xmajorgrids,
xmin=0, xmax=5,
xtick style={color=black},
y grid style={white!69.0196078431373!black},
ylabel={PSNR [dB]},
xlabel={BPP [Bytes]},
ymajorgrids,
ymin=0, ymax=140,
			]

\addplot [color=mycolor7!40!green, line width=1.0pt, mark size=3pt, mark=o, mark options={solid, fill=mycolor7!40!green, mycolor7!40!green},only marks]
  table[row sep=crcr]{%
1.94573881 91	\\
};
\addlegendentry{Octree (LOW)}

\addplot [color=mycolor7!40!green, line width=1.0pt, mark size=3pt, mark=*, mark options={solid, fill=mycolor7!40!green, mycolor7!40!green},only marks]
  table[row sep=crcr]{%
4.45183148 124	\\
};
\addlegendentry{Octree (HIGH)}

\addplot [color=mycolor11, line width=1.0pt, mark size=3pt, mark=*, mark options={solid, fill=mycolor11, mycolor11},only marks]
  table[row sep=crcr]{%
0.54873 49.5  \\
};
\addlegendentry{G-PCC}

\addplot [color=color1, line width=1.0pt, mark size=2.5pt, mark=square*, mark options={solid, fill=color1, color1},only marks]
  table[row sep=crcr]{%
1.68421196 109\\
};
\addlegendentry{J-LS (Cartesian)}

\addplot [color=color1, line width=1.0pt,  mark size=2.5pt, mark=square, mark options={solid, fill=color1, color1},only marks]
  table[row sep=crcr]{%
1.61719068 111\\
};
\addlegendentry{J-LS (Spherical)}

\addplot [color=mycolor9!60!black, line width=1.0pt, mark size=3pt, mark=diamond*, mark options={solid, fill=mycolor9!60!black, mycolor9!60!black},only marks]
  table[row sep=crcr]{%
1.61029151 109\\
};
\addlegendentry{PNG (Cartesian)}

\addplot [color=mycolor9!60!black, line width=1.0pt,  mark size=3pt, mark=diamond, mark options={solid, fill=mycolor9!60!black, mycolor9!60!black},only marks]
  table[row sep=crcr]{%
1.45559919 111\\
};
\addlegendentry{PNG (Spherical)}

\addplot [color=mycolor9!40!red, line width=1.0pt,  mark size=3pt, mark=triangle*, mark options={solid, fill=mycolor9!60!red, mycolor9!60!red},only marks]
  table[row sep=crcr]{%
0.73 62.45\\
};
\addlegendentry{LZW (Spherical)}

\addplot [color=mycolor6, line width=1.0pt,  mark size=3pt, mark=triangle*, mark options={solid, fill=mycolor6, mycolor6},only marks]
  table[row sep=crcr]{%
0.92 49.97\\
};
\addlegendentry{MJ2 (Spherical)}




		\end{axis}

\draw[dashed,fill=yellow,fill opacity=0.2] (2.35,4.2) circle [radius=0.4];	
\node at (2.35,4.9) {\scriptsize working points};

\draw[dashed] (6.5,4.7) circle [radius=0.3];	
\node at (6.5,4) {\scriptsize $+14\%$ PSNR};
\node at (6.5,3.68) {\textcolor{red}{\scriptsize $+207\%$ BPP}};

\draw[dashed] (2.85,3.47) circle [radius=0.25];	
\node at (2.85,2.9) {\textcolor{red}{\scriptsize $-22\%$ PSNR}};
\node at (2.85,2.6) {\textcolor{red}{\scriptsize $+33\%$ BPP}};

\draw[dashed] (1.3,2.18) circle [radius=0.4];	
\node at (2.3,1.9) {\textcolor{red}{\scriptsize $-45\%$ PSNR}};
\node at (2.3,1.62) {{\scriptsize $-45\%$ BPP}};

\draw[dashed] (.8, 1.88) circle [radius=0.2]; 
\node at (0.7,1.5) {\textcolor{red}{\scriptsize $-55\%$ PSNR}};
\node at (0.7,1.2) {{\scriptsize $-66\%$ BPP}};
		
\end{tikzpicture}
    \vspace{-0.43cm}
    \caption{Compression quality (in terms of PSNR) vs. efficiency (in terms of BPP) trade-off for different 2D vs. 3D compression methods.}
    \label{fig:comparison}
  \end{figure}
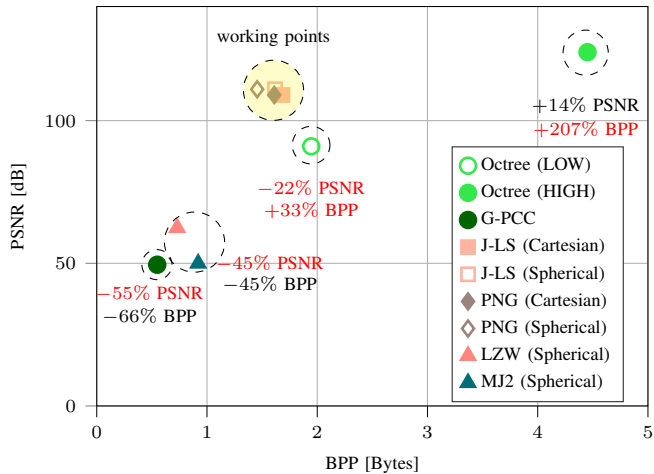

\textbf{Compression guidelines for data broadcasting.}
To summarize our conclusions, Fig.~\ref{fig:comparison} compares the compression performance of the investigated algorithms in terms of PNSR (to quantify the accuracy of the reconstructed point cloud) and BPP (to quantify the size of the compressed point cloud).
As anticipated, image-based methods, in particular PNG, achieve the best trade-off.
On one side, Octree-based solutions at HIGH profile could guarantee up to $+14\%$ better PSNR, while requiring in turn $3\times$ more BPPs for compression, making this solution ineffective for efficient data broadcasting. A LOW profile would exhibit worse PSNR and BPP performance, and involve up to $10\times$ slower compression.
On the other side, video-based compression methods and G-PCC permit to represent the point cloud with $-45\%$ and $-66\%$ BPPs compared to PNG, even though degrading the PSNR by an impressive $-45\%$ and $-55\%$, which makes these solutions lossy. Also, MJ2, LZW and G-PCC are not compatible with low latency for data dissemination, since compression and decompression may take tens/hundreds of seconds to complete.
For both PNG and J-LS schemes, employing spherical coordinates before compression guarantees better compressibility than Cartesian-based methods, despite involving slower compression.





\section{Conclusions and Future Work} 
\label{sec:conclusions_and_future_work}
In this paper we faced the challenge of compressing LiDAR data to facilitate efficient data broadcasting. To do so, we compared 3D compression methods (Octrees and G-PCC) specifically designed for point clouds, and 2D methods (PNG, J-LS, LWZ and MJ2) typically used to compress image and video frames. 
We showed that 2D algorithms, even though requiring the raw point cloud to be first transformed into its two-dimensional representation, can achieve a high compression rate and up to $20\times$ faster compression than G-PCC, while guaranteeing a PNSR greater than 100 dB, thus supporting lossless compression. 

In our future work we will investigate whether more advanced solutions, e.g., other settings of G-PCC~\cite{graziosi2020overview} or methods based on artificial intelligence, may improve compression accuracy by operating directly on the raw 3D point cloud. 
\section*{Acknowledgments}
This work was partially supported by MIUR (Italian Ministry for Education and Research) under the initiative ``Departments of Excellence'' (Law 232/2016).
Paolo Testolina's activities were also supported by Fondazione CaRiPaRo under the grants ``Dottorati di Ricerca'' 2019.

\bibliographystyle{IEEEtran}
\bibliography{./bibl.bib}

\end{document}